\input amstex
\documentstyle{amsppt}

\def\Jac{\operatorname {Jac}}
\def\diag{\operatorname {diag}}
\topmatter
\title The Lagrange bitop  on $so(4)\times so(4)$\\
 and geometry of the Prym varieties \endtitle
\author Vladimir Dragovi\' c and Borislav Gaji\' c \endauthor
\rightheadtext{Lagrange bitop}
\abstract 
A four-dimensional integrable rigid-body system is considered and it is shown that
it represents two twisted three-dimensional Lagrange tops.  A polynomial Lax representation, which doesn't fit neither in Dubrovin's nor in Adler-van Moerbeke's picture is presented. The algebro-geometric integration procedure is based on deep  facts from the geometry of the Prym varieties of double coverings
of hyperelliptic curves. The correspondence between all such coverings with Prym varieties splitted as a sum of two varieties of the same dimension and the integrable hierarchy associated to the initial system is established. 
\endabstract 
\endtopmatter

\

\

\centerline{addresses:}

\

\centerline {V. D. {\it International School for Advanced Studies, Via Beirut 2-4, Trieste, Italy}} 

\

\centerline {\it Mathematical Institute SANU, Kneza Mihaila 35, Belgrade, Yugoslavia}

\

\centerline {email {\it vladad\@ mi.sanu.ac.yu and dragovic\@ sissa.it}}

\

\

\centerline {B. G. {\it Mathematical Institute SANU, Kneza Mihaila 35, Belgrade, Yugoslavia}}

\

\centerline {email {\it gajab\@ mi.sanu.ac.yu}} 

\

\

\

\

\

\newpage

\centerline{\bf 1. Introduction}

\

Starting from the end of 60's, the Lax representation has been the one of the most 
powerful tools in the inverse scattering method for the integration of 
nonlinear differential equations, partial and ordinary as well. The Lax equation 

$$
\dot L (\lambda) =[L(\lambda), A(\lambda)]
$$
with $L(\lambda), A(\lambda)$ being matrix polynomials in  so called spectral 
parameter $\lambda$ were studied in the middle seventies by Dubrovin 
[12, 15]. That theory was based on the notion of the Baker-Akhiezer 
function, developed by Krichever and others from Novikov's school (see [20,
14]). The $L-A$ pairs of such
type and Dubrovin's theory were used in the algebro-geometric integration of 
rigid-body motion in [21, 9].

Few years later, Adler and van Moerbeke presented a new version of such a 
theory in [1]. It was based on [22] and also applied in the integration of 
rigid-body motion in several cases [1, 25, 26]

Recently we have found a Lax representation of that, polynomial, form for 
the classical Hess-Apel'rot system, see [11]. Generalizing it, we 
constructed a new completely integrable system of the classical Euler-Poisson equations
of motion of a heavy four-dimensional rigid body fixed at a point. Together
with generalized Lagrange case and generalized symmetric case, which were
introduced by Ratiu
(see [25]), this system exosted the list of integrable Euler-Poisson equations
with the $L$ operator, which is a quadratic polynomial in $\lambda$ of the form

$$
L(\lambda)=\lambda ^2C+\lambda M + \Gamma .
$$

The principal aim of this paper is to give 
algebro-geometric integration of this new system.

However, it appeared that this system did not fit exactly neither in Dubrovin's 
nor in Adler-van Moerbeke's picture. The matrix $L$ satisfies the condition
$$
L_{12}=L_{21}=L_{34}=L_{43}=0.
\tag 1
$$
Such situation is explicitly excluded by Adler-van Moerbeke (see [1], Theorem
1) and not so explicitly by Dubrovin (see [12], Lemma 5 and Corollary). Up to
our knowledge, examples which satisfy the condition of type (1) have not been
studied before.

Analysis of the spectral curve and the Baker-Akhiezer function shows that
dynamics of the system is related to certain Prym variety $\Pi$ (which splits
according to the Mumford-Dalalian theory [24, 10, 28]) and evolution of
divisors of some meromorphic differentials $\Omega^i_j$.Then the condition (1)
requires that differentials
$$
\Omega^1_2,\Omega^2_1,\Omega^3_4,\Omega^4_3
$$
have to be {\it holomorphic} during the whole evolution. Compatibility of this
requirement with dynamics poseses a strong constraint on the spectral curve:
{\it its theta divisor should contain some torus}. In the case presented here
such constraint appears to be satisfied according to Mumford's relation (see
 [24])
$$
\Pi^-\subset \Theta ,
\tag 2
$$
where $\Pi^-$ is a translation of the Prym variety $\Pi$.

The paper is organized as follows. In section 2 the definition of the system 
of the Euler-Poisson equations on $so(4)\times so(4)$ is given and few of its 
basic properties are listed such as the $L-A$ pair, a set of first integrals
in involution. In the section 3 transformation of coordinates is performed in
classical manner and the connection with the Lagrange top is presented. The
spectral curve is described in section 4. In section 5 the
Baker-Akhiezer function was studied. The next, section 6, contains analysis
of the Prym variety $\Pi$ and via the Mumford-Dalalian theory, the connection
of algebro-geometric and classical approach from section 3 is established.
In the section 7 differentials $\Omega^i_j$ are defined and the holomorphicity
condition is derived from the condition (1). By using Mumford's relation (2)
formulae in the theta functions were derived. Necessary gluing of the infinite
points of the spectral curve and passage to the generalized Jacobian is done
in the section 8. The whole hierarchy of the Lagrange bitop is considered
in the section 9. The higher operators $L_N$, are polynomials
of order $N$ in $\lambda$. Their spectral curves are double covering of
hyperelliptic curves of genus $2N-1$; these coverings are defined by those
divisors of order two which are of degree $2N$. We can conclude by saying that
the Lagrange bitop hierarchy realizes all coverings of that kind.

\ 

\

\centerline{\bf 2. The definition of the system and basic properties}

\

The equations of motion of a heavy $n$-dimensional rigid body fixed at a
point in the moving frame are:
$$
\aligned
\dot M&=\left[ M,\Omega \right] +\left[ \Gamma , \chi \right],\\
\dot \Gamma &=\left[ \Gamma , \Omega \right],
\endaligned
\tag 3
$$
where the moving frame is such that the matrix $I$ is diagonal in it,
$diag(I_1,\dots ,I_n)$. Here $M_{ij}=(I_i+I_j)\Omega _{ij}\in so(n)$ is the
kinetic momentum, $\Omega \in so(n)$ is the angular velocity, $\chi \in so(n)$
is a given constant matrix (describing  a generalized center of the mass),
$\Gamma \in so(n)$. Then $I_i + I_j$ are the principal inertia momenta. These
equations are on the semydirect product $so(n)\times  so(n)$ and they were 
introduced in [25].

We are going to consider a four-dimensional case of these equations defined by
$$
\aligned
I_1&=I_2=a\\
I_3&=I_4=b
\endaligned  \quad
\text{   and     }\quad
\chi =
\pmatrix 0 & \chi _{12} & 0 & 0 \\
-\chi _{12} & 0 & 0 & 0\\
0& 0 & 0 & \chi _{34}\\
0 & 0 & -\chi _{34} & 0
\endpmatrix
\tag 4
$$
with the conditions $a\ne b,\,\,\chi_{12}, \chi_{34}\ne0, |\chi_{12}|\ne |\chi_{34}|$.

We will call this system {\it the Lagrange bitop} for the reasons we will
explain at the end of section 3.

\

\proclaim{Proposition 1. [11]} The equations of motion (3) under the conditions
(4) have an $L-A$ pair representation
$$
\aligned
\frac d{dt} L(\lambda ) &=\left[ L(\lambda ), A(\lambda )\right] \\
L(\lambda )&=\lambda ^2 C+\lambda M +\Gamma \\
A(\lambda )&=\lambda \chi +\Omega,
\endaligned \tag 5
$$
where $C=(a+b)\chi $.
\endproclaim

\

Before analysing the spectral properties of the matrices $L(\lambda )$, we will
change the coordinates in order to diagonalize the matrix $C$.
In this new basis the matrices $L(\lambda)$ have the form
$\tilde L (\lambda )=U^{-1} L(\lambda ) U,$ where

$$
U=
\pmatrix
0 & 0 & \frac {i\sqrt 2}2 & \frac {\sqrt 2}2 \\
0 & 0 & \frac {\sqrt 2}2 & \frac {i\sqrt 2}2 \\
\frac {i\sqrt 2}2 & \frac {\sqrt 2}2 & 0 & 0 \\
\frac {\sqrt 2}2 & \frac {i\sqrt 2}2 & 0 & 0
\endpmatrix
$$
After strightforward calculations, we have

$$
\tilde L(\lambda )=
\pmatrix
-i\Delta _{34} & 0 & -\beta _3^{*} -i\beta _4^{*} & i\beta _3-\beta _4 \\
0 &  i\Delta _{34} &  -i\beta _3^{*} -\beta _4^{*} & -\beta _3+i\beta _4\\
\beta _3-i\beta _4 & -i\beta _3 +\beta _4 & -i\Delta _{12} & 0 \\
i\beta _3^{*} +\beta _4^{*} & \beta _3^{*}+i\beta _4^{*} & 0 & i\Delta _{12}
\endpmatrix \tag 6
$$
where
$$
\aligned
\Delta _{12}&=\lambda ^2 C_{12}+\lambda M_{12}+\Gamma _{12},\\
\Delta _{34}&=\lambda ^2C_{34}+\lambda M_{34}+\Gamma _{34},
\endaligned
$$
$$
\xalignat 2
\beta _3&=x_3+\lambda
y_3, & x_3&=\frac 12 \left( \Gamma _{13}+i\Gamma _{23}\right),\\
 \beta _4 &=
x_4+\lambda y_4, & x_4&=\frac 12 \left( \Gamma _{14}+i\Gamma _{24}\right),\tag 7\\
 \beta _3^{*}&=\bar x_3+\lambda \bar y_3, & y_3&=\frac 12 \left( M_{13}+
iM_{23}\right),\\
\beta _4^{*}&=\bar x_4+\lambda \bar y_4, & y_4&=\frac 12 \left( M_{14}+
iM_{24}\right).
\endxalignat
$$
The spectral polynomial 
$$
p(\lambda, \mu )=det \left ( \tilde L(\lambda )-\mu \cdot 1\right)
$$
has the form 
$$
p(\lambda , \mu )=\mu ^4+P(\lambda )\mu ^2 +[Q(\lambda )]^2,\tag 8
$$
where
$$
\aligned
P(\lambda)&=\Delta _{12}^2+\Delta _{34}^2+4\beta _3\beta _3
^{*}+4\beta _4\beta _4^{*},\\
Q(\lambda)&=\Delta _{12}\Delta _{34}+2i(\beta _3^{*}\beta _4 -
\beta _3\beta _4^{*}).
\endaligned
\tag 9
$$
We can rewrite it in terms of  $M_{ij}$ and $\Gamma _{ij}$: 
$$
\aligned
P(\lambda )&=A\lambda ^4 +B\lambda ^3+D\lambda ^2+E\lambda +F,\\
Q(\lambda )&=G\lambda ^4+H\lambda ^3+I\lambda ^2+J\lambda +K.
\endaligned \tag 10
$$
Their coefficients 
$$
\aligned
A&=C_{12}^2+C_{34}^2=\langle C_+,C_+\rangle +\langle C_-,C_- \rangle,\\
B&=2C_{34}M_{34}+2C_{12}M_{12}=2\left( \langle C_+,M_+\rangle +\langle C_-, 
M_-\rangle \right),\\
D&=M_{13}^2+M_{14}^2+M_{23}^2+M_{12}^2+M_{34}^2+2C_{12}\Gamma _{12}+2C_{34}
\Gamma _{34}\\
&=\langle M_+,M_+\rangle +\langle M_-,M_-\rangle +2\left( \langle C_+, 
\Gamma _+\rangle +\langle C_-,\Gamma _- \rangle \right), \\
E&=2\Gamma _{12}M_{12}+2\Gamma _{13}M_{13}+2\Gamma _{14}M_{14}+2\Gamma _{23}
M_{23}+2\Gamma _{24}M_{24}+2\Gamma _{34}M_{34}\\
&=2\left( \langle \Gamma _+, M_+ \rangle +\langle \Gamma _- , M_- \rangle
\right), \\
F&=\Gamma _{12}^2+\Gamma _{13}^2+\Gamma _{14}^2+\Gamma _{23}^2+\Gamma _{24}^2
+\Gamma _{34}^2=\langle \Gamma _+ , \Gamma _+ \rangle +\langle \Gamma _-, 
\Gamma _-\rangle ,\\
G&=C_{12}C_{34}=\langle C_+, C_-\rangle,\\
H&=C_{34}M_{12}+C_{12}M_{34}=\langle C_+,M_-\rangle +\langle C_-, M_+\rangle,\\
I&=C_{34}\Gamma _{12}+\Gamma _{34}C_{12}+M_{12}M_{34}+M_{23}M_{14}-M_{13}M_{24}\\
&=\langle C_+, \Gamma _- \rangle +\langle C_-,\Gamma _+\rangle +\langle M_+,
M_- \rangle , \\
J&=M_{34}\Gamma _{12}+M_{12}\Gamma _{34}+M_{14}\Gamma _{23}+M_{23}\Gamma _{14}-
\Gamma _{13}M_{24}-\Gamma _{24}M_{13}\\
&=\langle M_+, \Gamma _- \rangle +\langle M_-, \Gamma _+ \rangle, \\
K&=\Gamma _{34}\Gamma _{12}+\Gamma _{23}\Gamma _{14}-\Gamma _{13}\Gamma _{24}
=\langle \Gamma _+, \Gamma _-\rangle.
\endaligned \tag 11
$$
are integrals of motion of the system (3, 4). We used two vectors
$M_+,M_-\in R^3$ which correspond to  $M_{ij}\in so(4)$ according to 
$$
(M_+,M_-)\rightarrow
\pmatrix
0 & -M^3_{+} & M^2_{+} & -M^1_{-}\\
M^3_{+} & 0 & -M^1_{+} & -M^2_{-}\\
-M^2_{+} & M^1_{+} & 0 & -M^3_{-}\\
M^1_{-} & M^2_{-} & M^3_{-} & 0
\endpmatrix
\tag{12}
$$
Here $M_{+}^j$ are the $j$-th coordinates of the vector $M_+$.
The system (3, 4) is Hamiltonian with the Hamiltonian function
$$
{\Cal H}=\frac12(M_{13}\Omega_{13}+M_{14}\Omega_{14}+M_{23}\Omega_{23}+
M_{12}\Omega_{12}+M_{34}\Omega_{34})+\chi_{12}\Gamma _{12}+\chi_{34}\Gamma_{34}
$$
The algebra $so(4)\times so(4)$ is 12 dimensional. The general orbits of the
coadjoint action are 8 dimensional. According to [25], the Casimir functions
are coefficients of $\lambda^0, \lambda, \lambda^4$ in the polynomials 
$[det \tilde L(\lambda )]^{1/2}$ and $-\frac 12 Tr (\tilde L(\lambda ))^2$. 

Since
$$
\aligned
\left[ det \tilde L(\lambda )\right] ^{1/2} &=G\lambda ^4+H\lambda ^3 +I\lambda ^2 +
J\lambda +K,\\
-\frac 12 Tr\left(\tilde L(\lambda )\right)^2 &=A\lambda ^4+E\lambda +F,
\endaligned
$$
the Casimir functions are $J,K,E,F$. Nontrivial integrals of motion are $B,D,H,I$.
They are in involution.

Nontrivial integrals of motion are $B,D,H,I$ are independent in the case 
$\chi_{12}\ne \pm  \chi_{34}$. When $|\chi_{12}|=|\chi_{34}|$, then $2H=B$ or $2H=-B$ and there are 
only 3 independent integrals in involution

So we have

\

\proclaim{Proposition 2 [11]} For $|\chi_{12}|\ne |\chi_{34}|$, the system (3, 4) 
is completely integrable in the Liouville sense.
\endproclaim

\

There are two families of integrable Euler-Poisson equations introduced by 
Ratiu in [25]. {\it The generalized symmetric case} is defined by the conditions
$$
I_1=\dots =I_n,\quad \chi\  \text{arbitrary;}
$$
and {\it the generalized Lagrange case} which is defined by
$$
I_1=I_2=a,\ \  I_3=\dots =I_n=b,\ \  \chi_{ij} =0\ \text{if}\ (i,j) 
\notin \{(1,2),(2,1)\}.
$$
The system (3, 4) doesn't fall in any of those families and together with them
it makes the complete list of systems with the $L$ operator of the form

$$
L(\lambda)=\lambda ^2C+\lambda M + \Gamma .
$$

\

\proclaim{Proposition 3 [11]} If $\chi_{12}\ne 0$ then the Euler-Poisson equations
(3) could be written in the form (5) (with arbitrary $C$) if and only if the
equations (3) describe the generalized symmetric case, the generalized Lagrange
case or the Lagrange bitop.
\endproclaim

\

One can compare this with [24] (Theorem 15, ch. 53) and [16] (Example 2, p. 1451 ).
The Lagrange bitop can be embedded, on the other hand,  in  the Bolsinov construction
([24], Theorem 17, ch. 53).

\

\

\centerline{\bf 3. Classical integration and the Lagrange system}

\

Starting from the well- known decomposition  $so(4)=so(3)\oplus so(3)$,
let us introduce
$$
M_1=\frac 12 (M_++M_-)\qquad M_2=\frac 12 (M_+-M_-)
$$
(and similar for $\Omega, \Gamma, \chi$), where $M_+, M_-$ are defined with 
(12). Equations (3) become
$$
\aligned
\dot M_1&=2(M_1\times\Omega_1+\Gamma_1\times\chi_1)\qquad
\dot \Gamma_1=2(\Gamma_1\times\Omega_1)\\
\dot M_2&=2(M_2\times\Omega_2+\Gamma_2\times\chi_2)\qquad
\dot \Gamma_2=2(\Gamma_2\times\Omega_2)
\endaligned
\tag{13}
$$
and 
$$
\aligned
\chi_1&=(0,0,-\frac12 (\chi_{12}+\chi_{34}))\\
\chi_2&=(0,0,-\frac12 (\chi_{12}-\chi_{34}))
\endaligned
$$

 Since 
$$
M_+=I_+\Omega_+,\quad 
M_-=I_-\Omega_-,
$$ 
where $I_+=\diag(I_2+I_3, I_1+I_3, I_1+I_2),\ I_-=\diag(I_1+I_4, I_2+I_4, 
I_3+I_4)$, the connection between $M_1, M_2$ and $\Omega_1, \Omega_2$  is 
$$
\aligned
M_1&=\frac12[(I_++I_-)\Omega_1+(I_+-I_-)\Omega_2]\\
M_2&=\frac12[(I_+-I_-)\Omega_1+(I_++I_-)\Omega_2].
\endaligned
$$
Using (4), we have
$$
\aligned
M_1&=((a+b)\Omega_{(1)1}, (a+b)\Omega_{(1)2}, (a+b)\Omega_{(1)3}+ 
(a-b)\Omega_{(2)3})\\
M_2&=((a+b)\Omega_{(2)1}, (a+b)\Omega_{(2)2}, (a-b)\Omega_{(1)3}+ 
(a+b)\Omega_{(2)3}).
\endaligned
$$
We denoted with $\Omega_{(i)j}$ the $j$ component of the vector $\Omega_i$ and 
similar notation we use for $\Gamma$ and $\chi$.

If we denote $\Omega_1=(p_1, q_1, r_1),\ \Omega_2=(p_2, q_2, r_2)$, 
then the first group of the equations (13) becomes
$$
\aligned
&{\dot p}_1-mq_1r_2=-n_1\Gamma_{(1)2}\\
&{\dot q}_1+mp_1r_2=n_1\Gamma_{(1)1}\\
&(a+b){\dot r}_1+(a-b){\dot r}_2=0
\endaligned
\tag{14}
$$
and
$$
\aligned
&{\dot p}_2-mq_2r_1=-n_2\Gamma_{(2)2}\\
&{\dot q}_2+mp_2r_1=n_2\Gamma_{(2)1}\\
&(a-b){\dot r}_1+(a+b){\dot r}_2=0
\endaligned
\tag{15}
$$
where 
$$
m=-\frac{2(a-b)}{a+b},\qquad n_1=-\frac{2\chi_{(1)3}}{a+b},\qquad
n_2=-\frac{2\chi_{(2)3}}{a+b}.
$$
The integrals of motion are
$$
\aligned
&[(a+b)r_1+(a-b)r_2)]\chi_{(1)3}=f_{11}\\
&(a+b)^2(p_1^2+q_1^2)+[(a+b)r_1+(a-b)r_2]^2+2(a+b)\chi_{(1)3}\Gamma_{(1)3}
=f_{12}\\
&(a+b)p_1\Gamma_{(1)1}+(a+b)q_1\Gamma_{(1)2}+[(a+b)r_1+(a-b)r_2]\Gamma_{(1)3}
=f_{13}\\
&\Gamma_{(1)1}^2+\Gamma_{(1)2}^2+\Gamma_{(1)3}^2=1\\
&[(a-b)r_1+(a+b)r_2)]\chi_{(2)3}=f_{21}\\
&(a-b)^2(p_2^2+q_2^2)+[(a-b)r_1+(a+b)r_2]^2+2(a+b)\chi_{(2)3}\Gamma_{(2)3}
=f_{22}\\
&(a+b)p_2\Gamma_{(2)1}+(a+b)q_2\Gamma_{(2)2}+[(a-b)r_1+(a+b)r_2]\Gamma_{(2)3}
=f_{23}\\
&\Gamma_{(2)1}^2+\Gamma_{(2)2}^2+\Gamma_{(2)3}^2=1
\endaligned
\tag{16}
$$
Introducing $\rho_i, \sigma_i$, defined with $p_i=\rho_i\cos\sigma_i$, 
$q_i=\rho_i\sin\sigma_i$, using (15) and (16) after some calculations, we  
get 
$$
\aligned
&\rho_1^2{\dot\sigma}_1+mr_2\rho_1^2=n_1(\frac{f_{13}}{a+b}-\alpha_1
\Gamma_{(1)3})\\
&[(\rho_1^2)^{\cdot}]^2=4n_1^2\rho_1^2[1-\frac{1}{n_1^2}(a_1+\rho_1^2)^2]-
4n_1^2(\frac{f_{13}}{a+b}-\alpha_1a_1-\frac{\alpha_1}{n_1}\rho_1^2)^2\\
&\rho_2^2{\dot\sigma}_2+mr_1\rho_2^2=n_2(\frac{f_{23}}{a+b}-\alpha_2
\Gamma_{(2)3})\\
&[(\rho_2^2)^{\cdot}]^2=4n_2^2\rho_2^2[1-\frac{1}{n_2^2}(a_2+\rho_2^2)^2]-
4n_2^2(\frac{f_{23}}{a+b}-\alpha_2a_2-\frac{\alpha_1}{n_2}\rho_2^2)^2
\endaligned
\tag{17}
$$
where
$$
\aligned
\alpha_1&=\frac{(a+b)r_1+(a-b)r_2}{a+b}\quad
\alpha_2=\frac{(a+b)r_2+(a-b)r_1}{a+b}\\
a_i&=\frac{\alpha_i^2(a+b)^2-f_{i2}}{(a+b)^2}\quad
i=1,2
\endaligned
$$
Let us denote $u_1=\rho_1^2,\ u_2=\rho_2^2$. From (17) we have
$$
\aligned
{\dot u}_i^2&= P_i(u_i),\qquad i=1,2.
\endaligned
$$
where
$$
\aligned
P_i(u)&=-4u^3-4u^2B_i+4uC_i+D_i,\qquad i=1,2.
\endaligned
$$
and
$$
\aligned
B_i&=2a_i+\alpha_i^2,\\
C_i&=n_i^2-a_i^2- 4\frac{\alpha_i\chi_{(i)3}f_{i3}}{(a+b)^2}-2\alpha_i^2a_i,\\
D_i&=-4(\frac{2\chi_{(i)3}f_{i3}}{(a+b)^2}+\alpha_ia_i)^2,\qquad i=1,2.
\endaligned
\tag{18}
$$
From the previous relations, we have
$$
\int\frac{du_1}{\sqrt{P_1(u_1)}}=t,\quad
\int\frac{du_2}{\sqrt{P_2(u_2)}}=t.
$$
So,  the integration of the system (13) leads to the functions associated
with the elliptic curves $E_1, E_2$,
where $E_i=E_i(\alpha_i, a_i, \chi_{(i)3}, f_{i2}, f_{i3})$ are 
given with:
$$
E_i:  y^2 = P_i(u).
\tag {19}
$$
The equations (14) and (15) are very similar to those for the classical
Lagrange system (see [17]). However, the system (14, 15) doesn't split on
two independent Lagrangian systems, since the third equations in (14) and
(15) together give that $r_1$ and $r_2$ are constants. Also, in the
definition of each of the curves $E_1, E_2$ both those constants are involved.
That is the reason we refer to the system (3, 4) as  nonsplitted {\it Lagrange
bitop}. The formulae (17) are also very close to those for the Lagrange system.
Although the Lagrange system has a long history (starting from 1788) and
many important papers have been  written about it, still there are some
subtle questions and problems related to its integration (see [16, 5]). So,
for the further analysis of the Lagrange bitop we pass to the
algebro-geometric integration.

\

\

\centerline{\bf 4. The first steps in an algebro-geometric integration procedure}

\

The $L(\lambda)$ matrix (5) for the Lagrange bitop (3, 4) is a quadratic polynomial
in the spectral parameter $\lambda $ with matrix coefficients. The general
theories describing the isospectral deformations for polynomials with matrix
coefficients were developed by Dubrovin [12, 15] in the middle of 70's and
by Adler, van Moerbeke [1] few years later. Dubrovin's approach was based
on the Baker-Akhiezer function and it was applied in rigid body problems in
[21, 9]. The other approach was based on [22] and the connection
with rigid body problems was given in [1, 25, 26].

As it will be shown bellow, non of these two theories can be directly applied
in our case. So, we are going to make certain modifications, and then
we will integrate the system (3, 4). As usual in the algebro-geometric
integration, we consider the spectral curve
$$
\Gamma : \, det\left( \tilde L(\lambda )-\mu \cdot 1 \right) =0.
$$
By using (8, 9), we have
$$
\Gamma : \, \mu ^4+\mu ^2\left( \Delta _{12}^2+\Delta _{34}^2+4\beta _3\beta _3
^{*}+4\beta _4\beta _4^{*}\right) +\left[ \Delta _{12}\Delta _{34}+2i(\beta
_3^{*}\beta _4-\beta _3\beta _4^{*})\right]^2=0.\tag {20}
$$
There is an involution
$$
\sigma:(\lambda,\mu)\rightarrow (\lambda, -\mu)
$$
on the curve $\Gamma$, which corresponds to the skew symmetry of the matrix $L(\lambda)$. Denote the factor-curve by  $\Gamma_1=\Gamma/\sigma$.

\proclaim{Lemma 1}
The curve $\Gamma_1$ is a smooth hyperelliptic curve of the  
genus $g(\Gamma_1)=3$. The arithmetic genus of the curve $\Gamma $
is $g_a(\Gamma) =9$.
\endproclaim
\demo{Proof} The curve:
$$
\Gamma_1:\, u^2+P(\lambda)u+[Q(\lambda)]^2=0,
$$
is hyperelliptic, and its equation in the
canonical form is:
$$
u_1^2=\frac{[P(\lambda)]^2}4- [Q(\lambda)]^2,
\tag{21}
$$
where $u_1=u+P(\lambda)/2$.
Since $\frac{[P(\lambda)]^2}4-[Q(\lambda)]^2$ is a polynomial of the
degree 8,
 the genus of the curve $\Gamma_1$ is $g(\Gamma_1)=3$.
The curve $\Gamma$ is a double covering of $\Gamma_1$, and the ramification
divisor is of the degree 8. According to the  Riemann-Hurwitz formula, 
the arithmetic genus of $\Gamma$ is $g_a(\Gamma)=9$.
\enddemo

\proclaim{Lemma 2} The spectral curve $\Gamma$ has four ordinary double 
points $S_i, i=1,\dots , 4$. The genus of its normalization  $\tilde \Gamma$ 
is five.
\endproclaim
\demo {Proof} From the equations
$$
\frac{\partial p(\lambda, \mu)}{\partial \lambda}=0,\quad
\frac{\partial p(\lambda, \mu)}{\partial \mu}=0,
$$
the four ordinary double  points are $S_k=(\lambda_k, 0), k=1,..,4$, 
where $\lambda_k$ are zeroes of the polynomial $Q(\lambda)$. Thus, the  
genus of the curve $\tilde \Gamma$ is $g(\tilde \Gamma)=g_a(\Gamma)-4=5$.
\enddemo

\proclaim{Lemma 3} The singular points $S_i$ of the curve $\Gamma$ are fixed 
points of the involution $\sigma$. The involution $\sigma $ exchanges the two branches of 
$\Gamma$ at $S_i$. 
\endproclaim

\demo{Proof} Fixed points of the $\sigma$ are defined with $\mu=0$, thus $S_i$ 
are fixed points. Since their projections on $\Gamma_1$ are smooth points, the involution $\sigma$ exchanges the branches of $\Gamma$, which are given by the 
equation
$$
\mu^2=\frac{-P(\lambda)+\sqrt{P^2(\lambda)-4Q^2(\lambda)}}{2}.
$$

\enddemo

In general, whenever the  matrix $L(\lambda)$ is antisymmetric, the spectral curve
is reducible in odd-dimensional case and singular in even-dimensional
case.

Before starting the study of the analytic properties of the Baker-Akhiezer
function, let us give the formulae for (nonnormalized) eigen-vectors
of the matrix $L(\lambda)$.

\proclaim{Lemma 4} If the vector $f=(f_1,f_2,f_3,f_4)^t$ is given by

$$
\aligned
f_1&=(\Delta_{12}^2+\mu^2)(i\Delta_{34}-\mu)-
2\mu(\beta_3\beta^*_3+\beta_4\beta^*_4)
+2\Delta_{12}(\beta_3\beta^*_4-\beta_4\beta^*_3)\\
f_2&=2\mu(\beta_3-i\beta_4)(i\beta_3^*+\beta_4^*)\\
f_{3}&=(-\beta_3+i\beta_4)\left[(i\Delta_{12}-\mu)(i\Delta_{34}-\mu)
+2i(\beta_3\beta_4^*-\beta_3^*\beta_4)\right]\\
f_4&=(i\beta_3^*+\beta_4^*)\left[(i\Delta_{12}+\mu)(i\Delta_{34}-\mu)
+2i(\beta_3\beta_4^*-\beta_3^*\beta_4)\right]
\endaligned
\tag{22}
$$
then $L(\lambda)f=\mu f.$
\endproclaim

\proclaim {Corollary 1} The eigenvectors $f'$ normalized by the condition
$$
f_1'+f_2'+f_3'+f_4'=1,
$$
have different values in the points $S_i',S_i''\in \tilde \Gamma $,
which cover the singular points $S_i\in \Gamma$.
\endproclaim

\

\

\centerline{\bf 5. The Baker-Akhiezer function}

\

The general integration technique based on the Baker-Akhiezer function
was developed by Krichever (see [20], [14] and bibliography therein).
The application on the matrix polynomials was done, as we said, by Dubrovin
(see [12, 15]). Following those ideas, we consider the next eigen-problem

$$
\aligned
\left(\frac {\partial }{\partial t}+\tilde A(\lambda )\right) \psi _k&=0,\\
\tilde L(\lambda )\psi _k&=\mu _k \psi _k,
\endaligned \tag{23}
$$
where $\psi _k$ are the eigenvectors with the eigenvalue                
$\mu _k$. Then $\psi _k(t,\lambda )$ form $4\times 4$ matrix              
with components   $\psi _k^i(t,\lambda )$. Denote by  $\varphi _i^k$
corresponding inverse matrix. Let us introduce
$$
\bar{\Psi}_j^i(t,\tau, (\lambda , \mu _k))=\psi _k^i(t,\lambda )\cdot \varphi _j^k
(\tau,\lambda )
$$
and
$$
g_j^i(t,(\lambda , \mu _k))=\bar{\Psi}^i_j(t,t,(\lambda,\mu_k))
=\psi _k^i(t,\lambda )\cdot \varphi _j^k(t,\lambda )
$$
(there is no summation on $k$)
or, in other words
$$
g(t)=\psi _k(t)\otimes \varphi(t) ^k.
$$
It is easy to check that the function $\bar{\Psi}_i^j(t,\tau,(\lambda, \mu_k))$ 
satisfies
$$
\left(\frac{\partial}{\partial t}+\tilde{A}(\lambda)\right)
\bar{\Psi}(t,\tau,(\lambda, \mu_k))=0
$$
Then, if we denote with $\Phi(t,\lambda)$ the fundamental solution of 
$$
\left(\frac{\partial}{\partial t}+\tilde{A}(\lambda)\right)\Phi(t,\lambda)=0,
$$
normalized with $\Phi(\tau)=1$, there is a relation
$$
\bar{\Psi}(t,\tau,(\lambda,\mu_k))=\Phi(t,\lambda)g(\tau,(\lambda,\mu_k)).
$$
Matrix $g$ is of rank 1, and we have    
$$
\frac {\partial \psi}{\partial t}=-\tilde{A}\psi,\quad \frac {\partial \varphi }
{\partial t }=\varphi \tilde{A},\quad \frac {\partial g}{\partial t}=
[g,\tilde{A}].
$$
We can consider vector-functions $\psi _k(t,\lambda )=
\left(\psi^1_k(t, \lambda),...,\psi^4_k(t, \lambda)\right)^{T}$ as one vector-
function $\psi(t, (\lambda, \mu))=\left(\psi^1(t,(\lambda, \mu)),...,
\psi^4(t, (\lambda, \mu))\right)^{T}$
on the  curve  $\Gamma $ defined with $\psi ^i(t,(\lambda , \mu_k))=
\psi^i_k(t, \lambda)$.
The same we have for the matrix $\varphi ^k_i$. The relations
for the divisors of zeroes and poles of the functions $\psi ^i$ i $\varphi _i$
in the affine  part of the curve $\Gamma $ are: 

$$
\aligned
\left( g^i_j)\right)_a&=d_j(t)+d^j(t)-D_r-D'_s,
\endaligned \tag {24}
$$
where  $D_r$ is the ramification divisor over $\lambda$ plane(see [12]) and
$D_s$ is divisor of singular points, $D'_s\le D_s$. One
can easily calculate
$deg \,D_r=16, deg D_s=8$.

The matrix elements $g_j^i(t,(\lambda ,\mu _k))$  are meromorphic functions
on the curve $\Gamma $.
We need their asymptotics  in the neighborhoods of the points $P_k$, which
cover the point $\lambda =\infty $.
Let $\tilde \psi _k$ be the eigenvector of the matrix $\tilde L(\lambda )$
normalized in $P_k$ by the condition $\tilde \psi _k^k=1$, and let
$\tilde \varphi_i^k$ be the inverse matrix for $\tilde \psi _k^i$.
We will use the following

\

\proclaim{Lemma 5} The matrix elements of $g$ have another decomposition 
$$
g_j^i=\psi _k^i\varphi _j^k=\tilde \psi _k^i\tilde \varphi ^k_j.
$$
\endproclaim

\

The proof of the Lemma is an immediate consequence of the proportionality of
the vectors $ \psi _k$ and $\tilde \psi _k$ (and  $ \varphi^k$ and
$\tilde \varphi^k$).

\

\proclaim {Lemma 6} (a) The matrix $g$ has a representation
$$
  g=\frac {\mu ^3 + a_1\mu ^2 + a_2\mu + a_3}{\partial p(\lambda, \mu)/\partial \mu},
$$
  where
$$
a_1=L, a_2=PE+L^2, a_3=PL+L^3.
$$
(b) For the Lax matrix $L$ from (3) it holds
$$
       a_3= P(\lambda _i)L(\lambda _i)+L^3(\lambda_i)=0,
$$
for $\lambda _i: Q(\lambda_i)=0.$
\endproclaim

The proof of the Lemma follows from [12] and strightforward calculation. From
the (a) part one can see that the matrix $g$ could have poles at the singular
points of the spectral curve. But, from (b) we have

\

\proclaim {Corollary 2} The matrix $g$ doesn't have poles at the singular
points of the curve $\Gamma $.
\endproclaim

\

So, from now on, taking Corollaries 1 and 2 into account, we will consider
all the functions in this section as functions on the normalization
$\tilde \Gamma $ of the spectral curve $\Gamma $.

Since the functions $\tilde \psi _k^i$ i $\tilde \varphi _j^k$ are meromorphic
functions in the neighborhood of the points $P_k$, their asymptotics
can be calculated by expanding  $\tilde \psi _k$ as a power series in
$\frac 1{\lambda }$ in the neighborhood of the point $\lambda =\infty $
around the vector $e_k$, where $e_k^i=\delta _k^i$. We get     
$$
\aligned
&\left( \tilde C+\frac {\tilde M}{\lambda }+\frac {\tilde \Gamma }
{\lambda ^2}\right)\left( e_i+\frac {u_i}{\lambda }+\frac {v_i}{\lambda ^2}+
\frac {w_i}{\lambda ^3}+\dots \right)\\
&=\left( \tilde C_{ii}+\frac {b_i}{\lambda }+\frac {d_i}{\lambda ^2}+
\frac {h_i}{\lambda ^3}+\dots \right) \left( e_i+\frac {u_i}{\lambda }+
\frac {v_i}{\lambda ^2}+\frac {w_i}{\lambda ^3}+\dots \right),
\endaligned
$$
where the matrices $\tilde C, \tilde M$ and  $\tilde \Gamma $ are defined by
$$
\tilde L(\lambda )=\lambda ^2\tilde C+\lambda \tilde M+\tilde \Gamma.
$$
Comparing the same powers of $\frac 1{\lambda ^{\alpha }}$, we get the system
of the equations
$$
\aligned
&\tilde Ce_i=\tilde C_{ii}e_i\\
&\tilde Cu_i+\tilde Me_i=\tilde C_{ii}u_i+b_ie_i\\
&\tilde Cv_i+\tilde Mu_i+\tilde \Gamma e_i=\tilde C_{ii}v_i+b_iu_i+d_ie_i\\
&\tilde Cw_i+\tilde Mv_i+\tilde \Gamma u_i=\tilde C_{ii}w_i+b_iv_i+d_iu_i+
h_ie_i.
\endaligned \tag 25
$$
From the system (25), we have
$$
\aligned
(u_i)_i&=0,\quad (v_i)_i=0,\quad (w_i)_i=0\\
(u_i)_j&=\frac {\tilde M_{ji}}{\tilde C_{ii}-\tilde C_{jj}} \qquad j\ne i\\
(v_i)_j&=\frac 1{\tilde C_{ii}-\tilde C_{jj}}\left( \sum_{k\ne i}\frac
{\tilde M_{jk}\tilde M_{ki}}{\tilde C_{ii}-\tilde C_{kk}}-\frac {\tilde M_{ii}
\tilde M_{ji}}{\tilde C_{ii}-\tilde C_{jj}}+\tilde \Gamma _{ji}\right) \\
(w_i)_j&=\frac 1{C_i-C_j}\left[ \sum_{k\ne i}\tilde M_{jk}(v_i)_k+\sum_{k\ne i}
\tilde \Gamma _{jk}(u_i)_k-b_i(v_i)_j-d_i(u_i)_j\right]\\
b_i&=\tilde M_{ii}\\
d_i&=\sum_{k\ne i}\frac {\tilde M_{ik}\tilde M_{ki}}{\tilde C_{ii}-
\tilde C_{kk}}+\tilde\Gamma _{ii}\\
h_i&=\sum_{k\ne i}\tilde M_{ik}(v_i)_k+\sum_{k\ne i}
\tilde \Gamma _{jk}(u_i)_k
\endaligned \tag 26
$$
So the matrix  $\tilde \psi =\{ \tilde \psi _k^i\}$ in the neighborhood of
$\lambda =\infty $ has the form:
$$
\tilde \psi =1+\frac u{\lambda }+\frac v{\lambda ^2}+\frac w{\lambda ^3} +
O\left( \frac 1{\lambda ^3}\right).
\tag{27}
$$

Let the expansion of the matrix $\tilde \varphi =\{ \tilde \varphi _i^k \}$
be
$$
\tilde \varphi =1+\frac {u_1}{\lambda }+\frac {v_1}{\lambda ^2}+\frac {w_1}{\lambda ^3}
+O\left( \frac 1{\lambda ^3}\right).
$$
From $\tilde \psi \cdot \tilde \varphi =1$ we obtain
$$
\aligned
u_1&=-u\\
v_1&=u^2-v\\
w_1&=2uv-u^3-w.
\endaligned \tag {28}
$$
From $g=\tilde \psi _k\otimes \tilde \varphi ^k$ in the neighborhood of 
$\lambda =\infty $ it holds
$$
g_j^i(t,(\lambda ,\mu _k))=\tilde \psi _k^i(t,\lambda )\tilde \varphi _j^k(t,
\lambda ).
\tag {29}
$$
The last relation and (26) and (28) imply
$$
g_j^i(t,(\lambda ,\mu_k))=1 \quad \text{  for  } i=j=k.
$$
For $i=j\ne k $ and $i,k\in \{ 1,2\} $ or $i,k\in \{ 3,4\}$ we have
$$
g_j^i(t,(\lambda ,\mu_k))=\frac 1{\lambda ^4}v_{ik}(v_1)_{ki}+O\left(\frac 1
{\lambda ^5}\right).
$$
So the functions $g_j^i$ have a zero of the fourth order at the point $P_k$. 
Here we used the fact that $\tilde M_{12}=0, \tilde M_{34}=0$.

On the other hand, for $i=j\ne k$ and $i\in \{ 1,2\},\, k\in \{ 3,4\} $ or $k\in \{1,2\}$ and
$i\in \{ 3,4\}$ we have
$$
g_i^i(t,(\lambda ,\mu_k))=\frac 1{\lambda ^2}\frac {\tilde M_{ik}\tilde M_{ki}}
{(\tilde C_{kk}-\tilde C_{ii})^2}+O\left( \frac 1 {\lambda ^3}\right).
$$
Thus, the functions $g_i^i$ in this case have a zero of the second order at
$P_k.$

In the same manner, for $i\ne j,\, k\ne i,\, k\ne j,\, i,j\in \{1,2\}$ or
$i,j\in \{3,4\}$, it holds
$$
g_i^j(t,(\lambda ,\mu_k))=\frac 1{\lambda ^2}\frac {\tilde M_{ik}\tilde M_{kj}}
{(\tilde C_{kk}-\tilde C_{ii})(\tilde C_{kk}-\tilde C_{jj})}+O\left( \frac 1
{\lambda ^3}\right).
$$
So $g_j^i$ have a zero of the second order at $P_k.$ 

For $i\ne j,\, k=i,\, i,j\in \{ 1,2\}$ or $i,j\in \{ 3,4\}$ we have
$$
g_j^i(t,(\lambda ,\mu_k))=\frac {(v_1)_{ij}}{\lambda ^2}+O\left( 
\frac 1{\lambda ^3}\right)
$$
and $g_j^i$ have a zero of the second degree at $P_k$.

In the case $i\ne j,\, k=j,\, i,j\in \{ 1,2 \}$ or $i,j\in \{ 3,4\} $
the functions $g_j^i$ at $P_k$ also have a zero of the second degree.

For $i\ne j,\, k\ne i,\, k\ne j$ and $(i,j)\notin \{ (1,2),\, (2,1),\, (3,4),\,
(4,3)\}$ we have
$$
g_j^i(t,(\lambda ,\mu_k))=\frac {const}{\lambda ^3}+O\left( \frac 1{\lambda ^4}
\right)
$$
So, the degree of a zero at $P_k$ of the functions $g_i^j $ in this case is 3.

In the last two cases the functions  $g_j^i $ have simple zeroes at $P_k$:

for $i\ne j,\, k=j,\, (i,j)\notin \{ (1,2),\, (2,1),\, (3,4),\, (4,3)\}$ the
expansion is
$$
g_j^i(t,(\lambda ,\mu_i))=\frac 1{\lambda }\frac {\tilde M_{ji}}{\tilde C_{ii}-
\tilde C_{jj}}+O\left( \frac 1{\lambda^2 }\right)
\tag{30}
$$
and similarly for $i\ne j,\, k=i,\, (i,j)\notin \{ (1,2),\, (2,1),\, (3,4),\, (4,3)\}$
we have
$$
g_j^i(t,(\lambda ,\mu_j))=\frac 1{\lambda }\frac {\tilde M_{ji}}{\tilde C_{jj}-
\tilde C_{ii}}+O\left( \frac 1{\lambda^2 }\right)
$$

Summarizing, we get

\

\proclaim{Lemma 7} The functions  $g_j^i$ have the following divisors in
infinity:

for $i\ne j, \, (i,j)\notin \{ (1,2),\, (2,1),\, (3,4),\, (4,3)\}$
$$
(g_j^i)_{\lambda =\infty }=3\left( P_1+P_2+P_3+P_4\right) -2P_i-2P_j;
$$
for $i,j\in \{ 1,2\}$ or $i,j\in \{ 3,4\}$
$$
(g_j^i)_{\lambda =\infty }=2\left( P_1+P_2+P_3+P_4\right) \quad \text{ for  }
i\ne j,
$$
$$
(g_j^i)_{\lambda =\infty }=2\left( P_1+P_2+P_3+P_4\right) -2P_i+2P_l \quad
\text{ for  } i=j,
$$
where $l\ne i$ and $l,i\in \{ 1,2\}$ or $l,i\in \{3,4\}$.
\endproclaim

\

Let us denote by $\tilde d_j$ and by $\tilde d^i$ the following divisors:
$$
\aligned
&\tilde d_1=d_1+P_2,\quad \tilde d_2=d_2+P_1,\quad \tilde d_3=d_3+P_4,\quad
\tilde d_4 =d_4+P_3,\\
&\tilde d^1=d^1+P_2,\quad \tilde d^2=d^2+P_1,\quad \tilde d^3=d^3+P_4,\quad
\tilde d^4 =d^4+P_3.
\endaligned
$$
Finally, we have
\proclaim{Proposition 4}

\noindent a) The divisors of matrix elements of $g$ are
$$
\left( g_j^i\right)=\tilde d^i+\tilde d_j-D_r+2\left(
P_1+P_2+P_3+P_4\right)-P_i-P_j \tag{31}
$$
b) The divisors $ \tilde d_i, \tilde d^j$ are of the same degree
$$
deg\, \tilde d_i=\deg \, \tilde d^j=5.
$$
\endproclaim
\demo{Proof}
a) is a consequence of the previous lemma.

From $deg\, \tilde d_i=\deg \, \tilde d^j=deg\, d_i+1=deg\, d^j+1$ and 
$deg\, D_r=16$ using (16) we have b).
\enddemo

For the functions
$$
\hat{\psi}^i(t,\tau,(\lambda,\mu_k))=\frac{\bar{\Psi}_j^i
(t,\tau,(\lambda,\mu_k))}{\sum_s g^s_j(\tau,(\lambda,\mu_k))}
$$
we have
$$
\hat{\psi}^i(t,\tau,(\lambda,\mu_k))=\sum_s\Phi^i_s(t,\lambda)h^s(\tau,(\lambda,\mu_k))
\tag{32}
$$
where $h^s$ are the eigenvector of $L(\lambda)$ normalized by the condition
$$
\sum_s h^s(t,(\lambda,\mu_k))=1
$$
From (32) it follows that
$$
\hat{\psi}^i(t,\tau,(\lambda,\mu_k))=\sum_s\Phi^i_s(t,\lambda)
\frac{\psi^s_k(\tau,\lambda)}{\sum_l\psi^l_k(\tau,\lambda)}=
\frac{\psi^i_k(t,\lambda)}{\sum_l\psi^l_k(\tau,\lambda)}.
\tag{33}
$$
\proclaim{Proposition 5} The functions $\hat{\psi}^i$ satisfy the following 
properties
\item {a)} In the affine part of $\tilde \Gamma$ the function $\hat{\psi}^i$
has 4 time dependent zeroes which belong to the divisor $d^i(t)$ defined by
formula (24), and 8 time independent poles, e.q.
$$
\left(\hat{\psi}^i(t,\tau,(\lambda,\mu_k))\right)_a=d^i(t)-\bar{\Cal D},
\qquad \deg\bar{\Cal D}=8.
$$
\item {b)} At the points $P_k$, the functions $\hat{\psi}^i$ have essential 
singularities as follows:
$$
\hat{\psi}^i(t, \tau , (\lambda , \mu )) = exp \, \left[ -(t-\tau )
R_k\right] \hat{\alpha}^i(t,\tau,(\lambda,\mu))
$$
where $R_k$ are given with
$$
R_1=i\left(\frac{\chi_{34}}{z}+\omega_{34}\right), 
R_2=-R_1, 
R_3=i\left(\frac{\chi_{12}}{z}+\omega_{12}\right),  
R_4=-R_3  
$$

and $\hat{\alpha}^i$ are holomorphic in a neighborhood of $P_k$,
$$
\hat{\alpha}^i(\tau,\tau,(\lambda,\mu))=h^i(\tau,(\lambda,\mu))
$$
and 
$$
\hat{\alpha}^i(t,\tau,P_k)=\delta_i^k+v^i_k(t)z+O(z^2),
$$
with
$$
v^i_k=\frac{\tilde{M_{ki}}}{\tilde{C_{ii}}-\tilde{C_{kk}}}.
$$

\endproclaim
\demo {Proof} From (33) we see that $\hat{\psi}^i$ has $d^i(t)$ as a
divisor of zeroes in the affine part. Also from (32) it follows that 
poles of $\hat{\psi}^i$ are poles of $h^s$, and do not depend on 
time. The functions $h^s$ are meromorphic on $\tilde \Gamma$, and they have the 
same number of poles and zeroes.

From (25) and (26), we  have that

-$h^1$ has simple zeroes at $P_3$ and $P_4$, and the double zero at $P_2$; 

-$h^2$ has simple zeroes at $P_3$ and $P_4$, and the double zero at $P_1$;

-$h^3$ has simple zeroes at $P_1$ and $P_2$, and the double zero at $P_4$;

-$h^4$ has simple zeroes at $P_1$ and $P_2$, and the double zero at $P_3$.

As in [20, 14], it could be proved that the functions $h^s$ have divisors
of poles $\bar D, \deg \bar D = 8$. 
 This proves the rest of a).

From 
$$
\frac {\partial \ln \hat{\psi}^i(t,\tau ,(\lambda ,\mu _k))}{\partial t}=
\frac {\dot{ \hat{\psi}} ^i(t,(\lambda , \mu ))}
{\hat{\psi}^i(t, (\lambda ,\mu ))}=
\frac {\sum A^i_l(t,\lambda )\psi^l_k(t,\lambda  )}{\psi^i_k(t,\lambda  )}
$$
and asymptotics (26), (27) for $\hat{\psi}^i_k$  in a neighborhood of the points $P_k$, using  the proportionality of $\psi^k$ and $\hat \psi^k$, we obtain
$$
\hat \psi ^i (t,\tau ) =exp [(t-\tau)R_k]\hat \alpha ^i(t,\tau ).
$$
Starting from the expansion in a neighborhood of $R_k$
$$
\psi^i_k(t,\lambda)=e^{R_kt}(a^i_k+v^i_k(t)z+O(z^2)),\quad
\varphi_i^k(t,\lambda)=e^{-R_kt}(b_i^k+w_i^k(t)z+O(z^2)),
$$
using (30) we get $a^i_k=\delta ^i_k$ and
$$
v^i_k=\frac{\tilde{M_{ki}}}{\tilde{C_{ii}}-\tilde{C_{kk}}}.
$$
This  finish the proof of the proposition.
\enddemo

\proclaim{Lemma 8} The following relation takes place on the Jacobian
$\Jac(\tilde \Gamma)$:
$$
\Cal A(d^j(t)+\sigma d^j(t))=\Cal A(d^j(\tau)+\sigma d^j(\tau))
$$
where $\Cal A$ is the Abel map from the curve $\tilde \Gamma$ to 
$\Jac(\tilde \Gamma)$. 
\endproclaim
\demo {Proof} Let us introduce functions $\varphi^i(t,\tau,(\lambda,\mu))=
\frac{\psi^i(t,0,(\lambda,\mu))}{\psi^i(\tau,0,(\lambda,\mu))}$. Then 
$(\varphi^i)_a=d^i(t)-d^i(\tau)$.
Using the relation  
$$
\sigma(P_1)=P_2, \qquad \sigma(P_3)=P_4,
$$
and statement b) in the Proposition 5 we see that $\sigma\varphi^i\cdot 
\varphi^i$ are meromorphic functions (they do not have the essential 
singularities in $P_k$). Consequently, for their divisors of zeroes and poles, it 
holds:
$$
(\sigma\varphi^i\cdot \varphi^i(t, \tau, (\lambda,\mu)))=
d^j(t)+\sigma d^j(t)-d^j(\tau)-\sigma d^j(\tau).
$$
Applying the Abel theorem, we finish the proof.
\enddemo
From the previous Lemma we see that the vectors $\Cal A (d^i(t))$ belong to some
translation of the Prym variety $\Pi = Prym(\tilde \Gamma|\Gamma_1)$. More
details concerning the Prym varieties one can find in [24, 23, 19, 7, 29, 28, 3,
 6]. The natural question arises to compare
two twodimensional tori $\Pi$ and $E_1\times E_2$, where the elliptic curves $E_i$ are defined in (19). 

\

\

\centerline {\bf 6. Geometry of the Prym variety $\Pi $}

\

Together with the curve $\Gamma_1, $ one can consider
curves $C_1$ and $C_2$ defined by the equations
$$
C_1: v^2=s(\frac{P(\lambda)}2 + Q(\lambda))
\tag{34}
$$
$$
C_2: v^2=s(\frac{P(\lambda)}2 - Q(\lambda))
\tag{35}
$$
where $s$ is a constant to be fixed in the next Lemma.

\

\proclaim{Lemma 9} If $s=2/(a+b)^2$ in (34, 35) then the curves $C_i$
are such that
$$
E_i =Jac(C_i)\qquad i=1,2.
$$
\endproclaim
\demo{Proof} The curves $E_i$ from (19) can be represented in a canonical form
$$
y^2=4u^3-g_{2i}u-g_{3i},\quad i=1,2,
\tag{36}
$$
where
$$
g_{2i}=4(\frac{B_i^2}3 +C_i),\quad g_{3i}=4(\frac{2B_i^3}{m27}+\frac{B_iC_i}3-\frac{D_i}4).
$$
On the other hand, the equations for curves $C_i$ are
$$
y^2=s(a_0^{\pm}\lambda ^4+a_1^{\pm}\lambda ^3+a_2^{\pm}\lambda ^2+a_3^{\pm}\lambda + a_4^{\pm}),
$$
where
$$
\aligned
a_0^{\pm}&=\frac12(C_{12}\pm C_{34})^2,\\
a_1^{\pm}&=(C_{12}\pm C_{34})(M_{12}\pm M_{34}),\\
a_2^{\pm}&=((M_{12}\pm M_{34})^2+(M_{23}\pm M_{14})^2+(M_{13}\mp M_{24})^2)/2\\         &+(C_{12}\pm C_{34})(\Gamma_{12}\pm \Gamma_{34}),\\
a_3^{\pm}&=(M_{12}\pm M_{34})(\Gamma_{12}\pm \Gamma_{34})+(M_{23}\pm M_{14})(\Gamma_{23}\pm \Gamma_{14})\\
         &+(M_{13}\mp M_{24})(\Gamma_{13}\pm \Gamma_{24}),\\
a_4^{\pm}&=2.
\endaligned
$$
The sign $+$ corresponds to the curve $C_1$ and $-$ to $C_2$.

Using the fact that the Jacobian of the curve given with the equation
$$
y^2=a\lambda ^4+4b\lambda ^3+6c\lambda ^2+4d\lambda +e
$$
is a canonical curve of the form (36) with
$$
g_2=ae-4bd+3c^2,\quad  g_3=ace+2bcd-ad^2-eb^2-c^3,
$$
we finish the proof of the Lemma by strightforward calculation.

\enddemo

Since the curve $ \Gamma_1$ is hyperelliptic, in a study of the
Prym variety $\Pi$ the Mumford-Dalalian theory can be applied (see [28, 24, 10]).
Thus, using the previous Lemma, we come to

\proclaim {Theorem 1} The following relations take place:
\item {a} The Prymian $\Pi$ is isomorphic to the product of the curves $E_i$:
$$
\Pi = Jac (C_1)\times Jac (C_2).
$$
\item {b} The curve $\tilde \Gamma $ is the desingularization of $\Gamma_1\times (P^1) C_2=C_1 \times (P_1) \Gamma_1.$
\item {c} The canonical polarization divisor $\Xi$ of $\Pi$ satisfies
$$
\Xi=E_1\times \Theta_2 + \Theta _1\times E_2
$$
where $\Theta _i $ is the theta - divisor of $E_i$.
 
\endproclaim
\demo {Proof} The Prym variety $\Pi$ corresponds to the unramified double
covering $\pi : \tilde \Gamma \rightarrow \Gamma_1$. This covering is determined
by the divisor $D \in Jac_2 (\Gamma_1)$, such that $2D=(\mu)$. So
$$
	D = R_1 + R_2 + R_3 + R_4 - 2(\bar P_1 + \bar P_3)
$$
where $R_i= \pi (S_i)$ are the projections of the singular points on $\Gamma$
and $\bar P_i = \pi (P_i)$  are the  projections of the infinite points on $\Gamma $.

On the other hand, the double covering over $\Gamma_1$ defined by the curves
$C_1, C_2$ corresponds to the divisor  $D_1 \in Jac_2 (\Gamma_1)$
$$
	D_1 = X_1 + X_2 + X_3 + X_4 - 2(\bar P_1 + \bar P_3)
$$
where $X_i$ are the branch points on $\Gamma_1$ defined by $P/2-Q=0$.
Simple calculation shows that 
$$
(\frac{\mu + P/2}{\mu-Q})^2=\frac {P/2-Q}{-2\mu }
$$
holds on $\Gamma_1$. From the last relation it follows that
the divisors $D$ and $D_1$ are equivalent. 

The rest of the theorem now follows from the Mumford - Dalalian theory [10, 24,
28].
\enddemo

Theorem 1 explains the connection between the curves $E_1, E_2$ and the Prym
variety $\Pi$. Further analysis of properties of the Prym varieties necessary
for the understanding of dynamics of the Lagrange bitop will be done in the 
next section.

\

\

\centerline {\bf 7.  Isoholomorphisity condition, Mumford's relation and}
\centerline {\bf integration using the Baker-Akhiezer function}

\

According to the Proposition 5, the Baker - Akhiezer function $\Psi$ satisfies
usual conditions of normalized (n=)4-point function on the curve of genus
$g=5$ with the divisor $\bar \Cal D$ of degree $\deg \bar \Cal D= g+n-1=8$, see [14, 13]. And by
the general theory, it should determine all dynamics uniquely. The basic
question is {\it why is such dynamics compatible with the condition (1)?}
In other words, why is the evolution of divisors $\tilde d_i(t)$ such that
{\it all the time} $\tilde d_1$ contains $P_2$, $\tilde d_2$ contains $P_1$ and
so on. To answer this question, let us consider the differentials $\Omega^i_j$
$$
        \Omega^i_j=g_{ij}d\lambda,\quad  i,j=1,\dots , 4.
$$
It was proven by Dubrovin in the case of general position, that
$\Omega^i_j$ is a meromorphic differential having poles at $P_i$ and $P_j$,
with residuums $v^i_j$ and $-v^j_i$ respectively.

We have a simple

\proclaim {Lemma 10} The condition (1) is equivalent to
$$
v^1_2=v^2_1=v^3_4=v^4_3=0.
\tag {37}
$$
\endproclaim

From the Lemma 10 and Corollary 2, it follows

\proclaim{Proposition 6} The four differentials
 $$
 \Omega^1_2, \Omega^2_1, \Omega^3_4, \Omega^4_3
 $$
 are holomorphic during the whole evolution.
 \endproclaim

\

We can say that the condition (1) (together with the Corollary 2) implies
{\it isoholomorphicity}.
Let us recall the general formulae for $v$ from [13].
$$
v^i_j=\frac{\lambda_i \theta (A(P_i)-A(P_j)+tU+z_0)} {\lambda_j
\theta (tU+z_0) \epsilon (P_i,P_j)}, i\ne j,
\tag {38}
$$
where $U =\sum x^{(k)}U^{(k)}$ is certain linear combination of $b$ periods $U^{(i)}$ of the differentials of the second kinde
$\Omega^{(1)}_{P_i}$, which have pole of order two at $P_i$; $\lambda_i$ are
nonzero scalars, and 
$$
\epsilon (P_i,P_j)=\frac {\theta [\ni ](A(P_i-P_j))}{(-\partial _{U^{(i)}}\theta [\ni ](0))^{1/2} (-\partial _{U^{(j)}}\theta [\ni ](0))^{1/2})}.
$$
(Here $\ni $ is an arbitrary odd nondegenerate characteristics.)

\

\proclaim{Proposition 7} Holomorphicity of some of the differentials
$\Omega^i_j$ implies that the theta divisor of the spectral curve
contains some torus.
\endproclaim

\

In a case of spectral curve which is a double unramified covering
$$
\pi: \tilde \Gamma \rightarrow \Gamma _1;
$$
with $g(\Gamma _1)=g, \quad g(\tilde \Gamma )=2g-1$,
as we have here, it is really satisfied that the theta divisor contains a torus,
see [24]. More precisely, following [24], let us denote by $\Pi^-$ the set
$$
   \Pi^-=\left\{ L\in Pic^{2g-2} \tilde \Gamma | Nm L = K_{\Gamma 1}, h^0(L) ~
   \text {is odd} \right\},
$$
where $K_{\Gamma _1}$ is the canonical class of the curve $\Gamma _1$ and
$Nm: Pic \tilde \Gamma \rightarrow Pic \Gamma _1$ is the norm map, see [24, 28]
for details. For us,  it is crucial that $\Pi^-$ is a translate of the Prym variety
$\Pi$ and that Mumford's relation ([24]) holds
$$
\Pi^-\subset \Theta _{\tilde \Gamma}.
\tag {39}
$$
Let us denote
$$
U= i(\chi_{34} U^{(1)}-\chi_{34} U^{(2)}+ \chi_{12} U^{(3)}-\chi_{12} U^{(4)}),
\tag {40}
$$
where $U^{(i)}$ is the vector of $\tilde b$ periods of the differential of the second kinde $\Omega^{(1)}_{P_i}$, which is normalized by the condition that $\tilde a$ periods are zero. We suppose here that the cycles $\tilde a, \tilde b$ on the curve $\tilde \Gamma$ and $a, b$ on $\Gamma_1$ are chosen  to correspond to the involution $\sigma $ and the projection $\pi$, see [2, 28]:
 $$
\aligned
\sigma (\tilde a_k)&= \tilde a_{k+2}, \quad k=1,2;\\
\pi (\tilde a_0)&=a_0; \quad \pi(\tilde b_0)= 2 b_0.
\endaligned
\tag{41}
$$
The basis of normalized holomorphic differentials $[u_0,\dots ,u_5]$ on $\tilde \Gamma $ and $[v_0, v_1,v_2]$ on $\Gamma_1$  are chosen such that
$$
\aligned
\pi^*(v_0)&=u_0,\\
\pi^*(v_i)&=v_i+\sigma(v_i)=v_i+v_{i+2},\quad i=1,2.
\endaligned
\tag{42}
$$

Now we have

\

\proclaim{Theorem 2} If the vector $z_0$ in (38) corresponds to the translation of
the Prym variety $\Pi$ to $\Pi^-$, and the vector $U$ is defined by (40) than
the conditions (37) are satisfied.
\endproclaim

\demo{Proof} Proof follows from the relations (38) and (40) and fact that
$P_2=\sigma (P_1)$ and $P_4=\sigma (P_3)$.
\enddemo

\proclaim{Proposition 8} The explicit formula for $z_0$ is 
$$
z_0=\frac{1}{2}(\hat \tau_{00},\hat \tau_{01},\hat \tau_{02},\hat \tau_{01},\hat \tau_{02}),
$$
where
$$
\hat \tau_{0i}=\int_{\tilde b_0} u_i, \quad i=0, 1, 2.
$$
\endproclaim
The proof follows from [19], the Proposition 4.7. 

\

The formulae for scalars $\lambda_i$ from the formula (38) will be given in the next section.

\

\

\

\centerline {\bf 8. The evolution on the generalized Jacobian}

\

The evolution on the Jacobian of the spectral curve, as we considered $\Jac (\tilde \Gamma)$ in the Section 7,  gives the possibility to reconstruct the evolution of the Lax matrix $L(\lambda)$ only up to the conjugation by diagonal matrices. That was one of the limitations  of the Adler-van Moerbeke approach (see [1], Theorem 1). To overcome this problem, we are going to consider, following Dubrovin, the generalized Jacobian, obtained by gluing together the infinite 
points; in  the present case $P_1, P_2, P_3, P_4:$ 
$$
\Jac(\tilde \Gamma| \left \{ P_1, P_2, P_3, P_4\right \}).
$$
It can be understood as a set of classes of relative equivalence among 
the divisors on $\tilde \Gamma $ of certain degree. Two divisors of the same degree  $D_1$
and $D_2$ are called {\it equivalent relative to the points} $P_1, P_2, P_3, P_4 $,
if there exists a function $f$ meromorphic on $\tilde \Gamma $ such that
$(f)=D_1-D_2$ and $f(P_1)=f(P_2)=f(P_3)=f(P_4)$.

The generalized Abel map is defined with
$$
\tilde A(P)=(A(P),\lambda_1 (P),...,\lambda_4(P)),
$$
where $A(P)$ is the standard Abel map and
$$
\lambda_i(P)=exp\int_{P_0}^P\Omega_{P_iQ_0}, i=1,...,4.
$$
Here $\Omega_{P_iQ_0}$ denotes the normalized differential of the third kinde,
with poles at $P_i$ and at arbitrary fixed point $Q_0$. 

Then the generalized Abel theorem (see [19]) can be formulated as

\

\proclaim {Lemma 11 (the generalized Abel theorem)} The divisors $D_1$ and $D_2$
are equivalent relative to the points $P_1, P_2, P_3, P_4.$ if and only if
there exist integervalued vectors $N, M$ such that
$$
\aligned
A(D_1)&= A(D_2) + 2\pi N +BM,\\
\lambda_j (D_1)& = c \lambda_j (D_2)exp(M,A(D_2)), j=1,...,4
\endaligned
$$
where $c$ is some constant and $B$ is the period matrix of the curve $\tilde \Gamma$.
\endproclaim

The generalized Jacobi inverse problem can be formulated as the question of
finding, for given $z$, points $Q_1,\dots, Q_8$ such that
$$
\aligned
\sum_1^8A(Q_i) - \sum_2^4A(P_i)&=z+K,\\
\lambda_j=c exp\sum_{s=1}^8\int_{P_0}^{Q_s}\Omega _{P_jQ_0} + \kappa_j, j=1,...4,
\endaligned
$$
 where the constants $\kappa_j$ depend on  the curve $\tilde \Gamma$, the points $P_1, P_2, P_3, P_4$ and the choice of local parameters around them. 

We will denote by $Q_s$ the points which belong to the divisor $\bar \Cal D$ from
the Proposition 5, and by $E$ the Pryme- form from [19]. Then we have

\proclaim{Proposition 9} The scalars $\lambda_j$  from the formula (38) are
given with
$$
\lambda_j=\lambda_j^0 exp \sum_{k\ne j}ix^{(k)}\gamma_j^k,
$$
where
$$
 \lambda_j^0=c exp\sum_{s=1}^8\int_{P_0}^{Q_s}\Omega _{P_jQ_0} + \kappa_j,
$$
vector $\vec x= (x^{(1)},\dots, x^{(4)})$ is $t(\chi_{34},-\chi_{34},\chi_{12},-\chi_{12})$ and
$$
\gamma_i^j=\frac{d}{dk_j}ln E(P_i,P)|_{P=P_j}.
$$
\endproclaim

\
To give the formulae for the Baker-Akhiezer function, we need some notations.
Let
$$
\alpha ^j(\vec x)=exp[i\sum \tilde \gamma _m^jx^{(m)}]\frac {\theta (z_0)}{\theta (i\sum x^{(k)}U^{(k)} +z_0)},
$$
where 
$$
\tilde \gamma ^j_m=\int _{P_0}^{P_j}\Omega^{(1)}_{P_m}, \quad m\ne j,
$$
and $\tilde \gamma^m_m$ is defined by the expansion
$$
\int _{P_0}^{P}\Omega^{(1)}_{P_m}=-k_m+\tilde \gamma^m_m + O(k_m^{-1}),\quad P \rightarrow P_m.
$$
Denote 
$$
\phi ^j(\vec x, P)= \alpha ^j(\vec x) exp(-i\int _{P_0}^P\sum x^{(m)}\Omega^{(1)}_{P_m})\frac {\theta (A(P)-A(P_j)-i\sum x^{(k)}U^{(k)}-z_0)}{\theta (A(P)-A(P_j)-z_0)}.
$$
Now we can state
\proclaim{Proposition 10} The Baker-Akhiezer function is given by
$$
\psi ^j(\vec x, P)=\phi ^j(\vec x, P)\frac {\lambda_j^0\frac {\theta (A(P-P_j)-z_0)}{\epsilon (P, P_j)}}{\sum_{k=1}^4 \lambda _k^0 \frac {\theta (A(P-P_k)-z_0)}{\epsilon (P, P_k)}}, \quad j=1,\dots ,4.
$$
\endproclaim 

\

The proofs of the statements in this Section are standard from Dubrovin's approach.

\

Having established how  parameters of the formula (38) evolve, the reconstruction of the evolution of the  phase space variables follows immediately from the Proposition 5.  The generalized Liouville tori are four dimensional. Since two of the integrals
of the motion of the Lagrange bitop are linear (see (11)), according to the well known fact of Classical Mechanics ([31, 3]) those generalized tori have twodimensional affine part. The two-dimensional compact part of such a torus corresponds to the real part of the two-dimensional Prymian $\Pi$. The affine part corresponds
to the odd part of the affine part of the generalized Jacobian 
$$
\Jac (\tilde \Gamma| \left\{P_1, P_2, P_3, P_4\right\}) = \Jac(\tilde \Gamma)\times \Cal {C^*}^3. 
$$
From the Theorem 1c, it follows that the reduction of the formulae can be done
up to the elliptic theta functions on $E_i$ and exponential functions.

\

\

\centerline {\bf 9. The Lagrange bitop hierarchy and}
\centerline {\bf equallysplitted double hyperelliptic coverings}

\

According to the Mumford - Dalalian theory (see [10, 24, 28]), double unramifide coverings over  a
hyperelliptic curve $y^2=P_{2g+2}(x)$ of genus $g$ are in the correspondence with the divisions of the set of the zeroes of the polynomial $P_{2g+2}$ on two
disjoint nonempty subsets  with even number of elements. We will consider those
coverings  which correspond to the divisions on  subsets  with {\it equal number of elements} and we  can call them {\it equallysplitted}, since the Prym variety splits then as a sum of two varieties of equal dimension. 

Now, let us consider with the fixed operator $A$ from (5) the whole hierarchy of  systems defined by the Lax equations
$$
\dot L_B^{(N)}=[L_B^{(N)},A],
$$
where 
$$
L_B^{(N)}(\lambda)=\lambda^N B+\lambda^{N-1} M_1+\dots +M_N
$$
is a polynomial in $\lambda$ of degree $N\ge 2$, and the matrix $B$ is proportional to the matrix $\chi$: $B=d\chi$. From the Lax equations we get the system
$$
\aligned
\frac{\partial M_N}{\partial t}&=[M_N,\Omega],\\
\frac{\partial M_k}{\partial t}+[\chi, M_{k+1}]&=[M_k,\Omega],\\
 [\chi, M_1]&=[B,\Omega].
\endaligned
$$

Generalizing the situation  from the Section 4, we see that the spectral
curve $\Gamma_N$ is a singular curve of the form
$$
p_N(\lambda , \mu )=\mu ^4+P_N(\lambda )\mu ^2 +[Q_N(\lambda )]^2 =0,
$$
where the polynomials $P_N,Q_N$ have degree $\deg P_N=\deg Q_N=2N$. So, its
normalization is a double covering over the hyperelliptic curve 
$$
\mu_1^2=\frac{P_N^2(\lambda)}{4}-Q_N^2(\lambda)
$$
of genus $g_N=2N-1$. This covering corresponds to the division of the set of
zeroes on subsets of zeroes of the polynomials $P_N/2-Q_N$ and $P_N/2+Q_N$.
This is an equallysplitted covering under the assumption $|\chi_{12}|\ne |\chi_{34}|$ we fixed at the beginning. It is easy to see that all equallysplitted 
coverings can be realized in such a way. So we have

\

\proclaim {Theorem 3} The Lagrange bitop hierarchy realizes all equallysplited
coverings over the hyperelliptic curves of genus greater than two.
\endproclaim

\

\

{\bf Acknowledgement} One of the authors (V. D.) has a great pleasure to
thank Professor B. Dubrovin for stimulating discussions and Professor M. Narasimhan for helpful observations; his research was partially supported by
SISSA and MURST Project Geometry of Integrable Systems. The research of both authors was partially supported by the Serbian Ministry of Science and Technology projects.

\

\

\

\centerline {\bf References}

\

\item {1} M. Adler and P. van Moerbeke. Linearization of Hamiltonian 
Systems, Jacobi Varieties and Representation Theory. {\it Advances in Math.} 
{\bf 38} (1980), 318-379.

\item{2} E. Arbarello, M. Cornalba, P. A. Griffiths and J. Haris. {\it 
Geometry of algebraic curves} (Springer-Verlag, 1985).

\item{3} V. I. Arnol'd. {\it Mathematical methods of classical mechanics}
(Moscow: Nauka, 1989 [in Russian, 3-rd edition]). 

\item{4} V. I. Arnol'd , V. V. Kozlov and A. I. Neishtadt. {\it Mathematical 
aspects of classical and celestial mechanics/ in Dynamical systems III} 
(Berlin: Springer-Verlag, 1988).

\item {5} M. Audin. {\it Spinning Tops} (Cambridge studies in Advanced 
Mathematics 51, 1996). 

\item{6} A. Beauville. Prym varieties and Schottky problem. {\it 
Inventiones Math.} {\bf 41} (1977), 149-196.

\item{7} E. D. Belokolos, A. I. Bobenko, V. Z. Enol'skii, A. R. Its and 
V. B. Matveev. {\it Algebro-geometric approach to nonlinear integrable 
equations} (Springer series in Nonlinear dynamics, 1994).

\item {8} A. I. Bobenko, A. G. Reyman, M. A. Semenov-Tien-Shansky.
The Kowalewski top 99 years later {\it Commun. Math. Phys.} {\bf 122} (1989),
321-354

\item {9} O. I. Bogoyavlensky Integrable Euler equations on Li algebras arising in physical problems {\it Soviet Acad Izvestya} {\bf 48} (1984), 883-938
[in Russian]

\item {10} S. G. Dalalian. Prym varieties of unramified double coverings of
the hyperelliptic curves {\it Uspekhi Math. Naukh} {\bf 29} (1974) 165-166
[in Russian]

\item {11} V. Dragovi\' c, B. Gaji\' c: An L-A pair for the Hess-Apel'rot
system and a new integrable case for the Euler-Poisson equations on $so(4)
\times so(4)$. {\it Roy. Soc. of Edinburgh: Proc A} {\bf 131} (2001), 845-855 

\item {12} B. A. Dubrovin. Vpolne integriruemye gamil'tonovy sistemy
svyazanye s matrichnymi operatorami i Abelevy mnogoobraziya. {\it Funk. 
Analiz i ego prilozhe\-niya} {\bf 11}, (1977 [in Russian]), 28-41.

\item {13} B. A. Dubrovin. Theta-functions and nonlinear equations.
{\it Uspekhi Math. Nauk}. {\bf 36} (1981 [in Russian]), 11-80.

\item {14} B. A. Dubrovin, I. M. Krichever and S. P. Novikov. Integrable
systems I. {\it in Dynamical systems IV}, (Berlin: Springer-Verlag, ) 173-280. 

\item {15} B. A. Dubrovin, V. B. Matveev, S. P. Novikov. Nonlinear
equations of Kortever-de Fries type, finite zone linear operators and 
Abelian varieties. {\it Uspekhi Math. Nauk}. {\bf 31} (1976 [in Russian]), 
55-136.

\item {16} L. Gavrilov, A. Zhivkov. The complex geometry of Lagrange top. 
{\it L'Ensei\-gnement Math\' ematique}. {\bf 44} (1998), 133-170

\item{17} V. V. Golubev. {\it Lectures on integration of the equations of
motion of a rigid body about a fixed point} (Moskow: Gostenhizdat, 1953 
 [in Russian]; English translation: Philadelphia: Coronet Books, 1953).

\item{18} P. A. Griffiths. Linearizing flows and a cohomological 
interpretation of Lax equations. {\it American Journal of Math} {\bf 107} 
(1983), 1445-1483.

\item {19} J. D. Fay. {\it Theta functions on Riemann surfaces}, Lecture Notes 
in Mathematics, vol. 352, (Springer-Verlag), (1973)

\item{20} I. M. Krichever. Algebro-geometric methods in the theory of
nonlinear equations {\it Uspekhi Math. Naukh} {\bf 32} (1977) 183 - 208

\item{21} S. V. Manakov. Remarks on the integrals of the Euler equations 
of the $n$-dimensional heavy top. {\it Funkc. Anal. Appl.} {\bf 10} (1976 
[in Russian]), 93-94.

\item {22} P. van Moerbeke and D. Mumford. The spectrum of difference 
operators and algebraic curves. {\it Acta Math.} {\bf 143} (1979), 93-154.

\item {23} D. Mumford. Theta characteristics of an algebraic curve. {\it Ann.
scient. Ec. Norm. Sup. 4 serie} {\bf 4}(1971), 181-192.

\item {24} D. Mumford. Prym varieties 1. {\it A collection of papers dedicated to
Lipman Bers} (Acad. Press.) New York, (1974), p. 325-350.

\item{25} T. Ratiu. Euler-Poisson equation on Lie algebras and the 
N-dimensional heavy rigid body. {\it American Journal of Math} {\bf 104} 
(1982), 409-448.

\item {26} T. Ratiu and P. van Moerbeke. The Lagrange rigid body motion.
{\it Ann. Ins. Fourier, Grenoble} {\bf 32} (1982), 211-234.

\item {27} A. G. Reyman and M. A. Semenov-Tian-Shansky: Lax representation 
with spectral parameter for Kovalevskaya top and its generalizations.
{\it Funkc. Anal. Appl.} {\bf 22} (1988 [in Russian]), 87-88.

\item {28} V. V. Shokurov. Algebraic curves and their Jacobians. {\it in
Algebraic Geometry III}, (Berlin: Springer-Verlag, 1998 ) 219-261.

\item {29} V. V. Shokurov. Distinguishing Prymians from Jacobians. {\it
Invent. Math.} {\bf 65} (1981) 209-219

\item{30} V. V. Trofimov and A. T. Fomenko. {\it Algebra and geometry of integrable 
Hamiltonian differential equations } (Moscow: Faktorial, 1995 [in Russian]).

\item {31} Whittaker {\it A treatise on the analytical dynamics of particles 
and rigid bodies}, Cambridge at the University Press, (1952), p.456

\end

\enddocument